\documentclass[a4paper,fleqn,usenatbib]{mnras}

\usepackage[T1]{fontenc}
\usepackage{ae,aecompl}

\usepackage{graphicx}	% Including figure files
\usepackage{amsmath}	% Advanced maths commands
\usepackage{amssymb}	% Extra maths symbols

\newcommand\simgt{\lower.5ex\hbox{$\; \buildrel > \over \sim \;$}}
\newcommand\simlt{\lower.5ex\hbox{$\; \buildrel < \over \sim \;$}}

\newcommand\rbb{R_{\rm b,2}}
\newcommand\lw{L_{\rm w}}
\newcommand\he{H_{\rm eff}}
\newcommand\hee{H_{\rm eff,2}}

\newcommand\msun{\rm{M_{\odot}}}

\usepackage[usenames,dvipsnames]{color}

\title[Accretion and dilution of pristine gas by early GCs]
{Accretion of pristine gas and dilution during the formation of multiple-population 
globular clusters}
\author [A. D'Ercole, F. D'Antona and E. Vesperini]
{A. D'Ercole$^{1}$\thanks{E-mail: annibale.dercole@oabo.inaf.it},
F. D'Antona$^{2}$, E. Vesperini$^{3}$\\
$^{1}$INAF, Osservatorio Astronomico di Bologna, via Ranzani 1, I-40127 Bologna, Italy\\
$^{2}$INAF, Osservatorio Astronomico di Roma, Via Frascati 33, I-00040 Monteporzio Catone (Roma), Italy\\
$^{3}$Department of Astronomy, Indiana University, Swain West, 727 E. 3rd Street, IN 47405 Bloomington (USA)
}

\date{Accepted XXX. Received YYY; in original form ZZZ}

\pubyear{2016}

\begin{document}
\label{firstpage}
\pagerange{\pageref{firstpage}--\pageref{lastpage}}
\maketitle

\begin{abstract}
  We study the interaction of the early spherical GC wind powered by
  Type II supernovae (SNe~II) with the surrounding ambient medium
  consisting of the gaseous disk of a star forming galaxy at redshift
  $z\simgt 2$. The bubble formed by the wind eventually breaks out of
  the disk, and most of the wind moves directly out of the galaxy and
  is definitively lost. The fraction of the wind moving nearly
  parallel to the galactic plane carves a hole in the disk which will
  contract after the end of the SN activity. During the interval of
  time between the end of the SN explosions and the ``closure" of the
  hole, very O-poor stars (the Extreme population) can form out of the
  super--AGB (asymptotic giant branch) ejecta collected in the GC
  center. Once the hole contracts, the AGB ejecta mix with the
  pristine gas, allowing the formation of stars with an oxygen
  abundance intermediate between that of the very O-poor stars and
  that of the pristine gas.  We show that this mechanism may explain
  why Extreme populations are present only in massive clusters, and can
  also produce a correlation between the spread in helium and the
  cluster mass.

  Finally, we also explore the possibility that our proposed mechanism
  can be extended to the case of multiple populations showing
  bimodality in the iron content, with the presence of two populations
  characterized by a small difference in [Fe/H].  Such a result
  can be obtained taking into account the contribution of delayed SN
  II.

\end{abstract}


%%%%%%%%%%%%%%%%%%%%%%%%%%%%%%%%%%%%%%%%%%%%%%%%%%

%%%%%%%%%%%%%%%%% BODY OF PAPER %%%%%%%%%%%%%%%%%%

\section{Introduction} 
\label{sec:introduction}
Many spectroscopic, spectrophotometric and photometric
observational studies have revealed that globular clusters (GCs) host multiple
stellar populations characterized by different chemical properties.
While the first stellar generation (FG) has a
chemical composition equal to that of the pristine gas, the second
generation (SG) is depleted in C and O, enhanced in N and Na, and in
some cases is also characterized by significant differences in helium;
with few exceptions, the iron abundance is the same in all the
stellar populations, indicating that the ejecta of the Type II
supernovae (SNe~II) belonging to the FG has been lost by the GC before
the formation of the stellar SG \citep[][and references
therein]{piotto15} {  which otherwise should be very likely 
Fe-enhanced, at odd with what observed in most of the clusters.}

In order to explain the picture provided by observational studies, a
number of theoretical frameworks differing in the FG sources of the
processed material out of which the SG forms have ben proposed: 
fast-rotating massive stars \citep{krause13}, massive interacting binaries \citep
{demink09,bastian13a}, asymptotic giant branch (AGB) and super-AGB
stars \citep{der08,der10}. Despite the numerous theoretical efforts,
many questions remain unanswered as none of the proposed scenarios is
exempt from problems and some arbitrary assumptions.r

All the proposed models require the processed ejecta of the FG sources
to be mixed and diluted with gas with pristine chemical composition.
In this paper we focus on the dynamics of the dilution in the
context of the AGB scenario, which has been originally proposed 
by \cite{der08} to account for the three populations, characterized by 
three different  helium abundances, discovered in NGC\,2808.
According to this scenario, the GC is initially
composed by FG stars with the chemical abundances of
the pristine gas out of which they formed; 
subsequently, the SN II
explosions clear the cluster of the remaining gas, and the super-AGB
and AGB ejecta can collect in the cluster centre through a cooling
flow; the stars created within this cooling flow form the SG
population.  In order to reproduce the observed Na-O anticorrelation
it is however necessary to mix pristine gas with the AGB ejecta
\citep[see][]{der10,der11,der12}. In this framework, the only stars
forming from pure AGB ejecta would be the very O-poor stars classified
as Extreme (E) by \citet{carretta10} while stars with less extreme O
depletion (those classified as Intermediate (I) by \citet{carretta10}
) would form from a mix of ejecta and pristine gas. For clusters
hosting E SG stars, dilution should start after the beginning of the
formation of the SG while in clusters where the E population is absent
SG formation should involve both AGB ejecta and pristine gas.

\citet{Renzini15} have suggested that AGB models are far from having
explored the entire parameter space, and a possible revision of the
cross section of the reaction rate destroying sodium could lead to a
reassessment of the need for the dilution. {  However, as already
  argued in \citet{der11}, dilution with pristine gas is
  likely to remain as an essential ingredient also in the case a
  revision of AGB models were to lead to yields characterized by a
  Na-O anticorrelation.}

Understanding the origin of this pristine gas and its dynamics during
the early evolutionary stages of globular clusters and the phase of SG
star formation is a key aspect in the study of multiple-population
globular cluster formation.  In this paper we present a model aimed at
following the evolution of the gas within a globular cluster forming
in the disk of a high-redshift galaxy.  We are first guided by the
idea of quantifying under which hypothesis the cluster gas may satisfy
the simplest constraints for the formation of multiple populations,
that is: $i$) initially the pristine gas must be completely removed,
at least from the clusters displaying an extended Na-O anticorrelation
(see above); $ii$) the cluster should be able to reaccrete the
pristine gas about 20--30 Myr after the end of the SN II
explosions\footnote{This time interval derives from the chemical
  models of \citet{der10,der12}.}; $iii$) the pristine gas should not
be contaminated by the SN II ejecta because in most clusters the
different stellar populations have the same iron abundances

We find out that a model fulfilling these requirements is compatible
with the formation of clusters in the disks of galaxies at $z>2$
\citep[see e.g.][and references therein]{kravtsov05,kruijssen15} and
may also explain why extreme populations are present only in massive
clusters.  Interestingly, the model may be easily extended to the case
of multiple populations showing bimodality in the iron content, with
the presence of two (or more) populations differing in [Fe/H] by a
generally small fraction. While the effects of the explosions of type
II supernovae `isolated in time' cannot be distinguished from the iron content 
of the clusters, which will be scarcely increased \citep{dantona16},
we show that there may be cases in which delayed-SNII
events may stop the cooling flow for a period of some decades of
million years, but their energy output is not able to push the ejecta much
beyond the cluster, allowing, at the end of this epoch, a new burst of
star formation to occur in gas polluted by both supernovae ejecta and
AGB ejecta.

\section{Set-up of the model}
\label{sec:mod}

\begin{figure*}    
\vskip -10pt
\centering{
 \includegraphics[width=17.5cm]{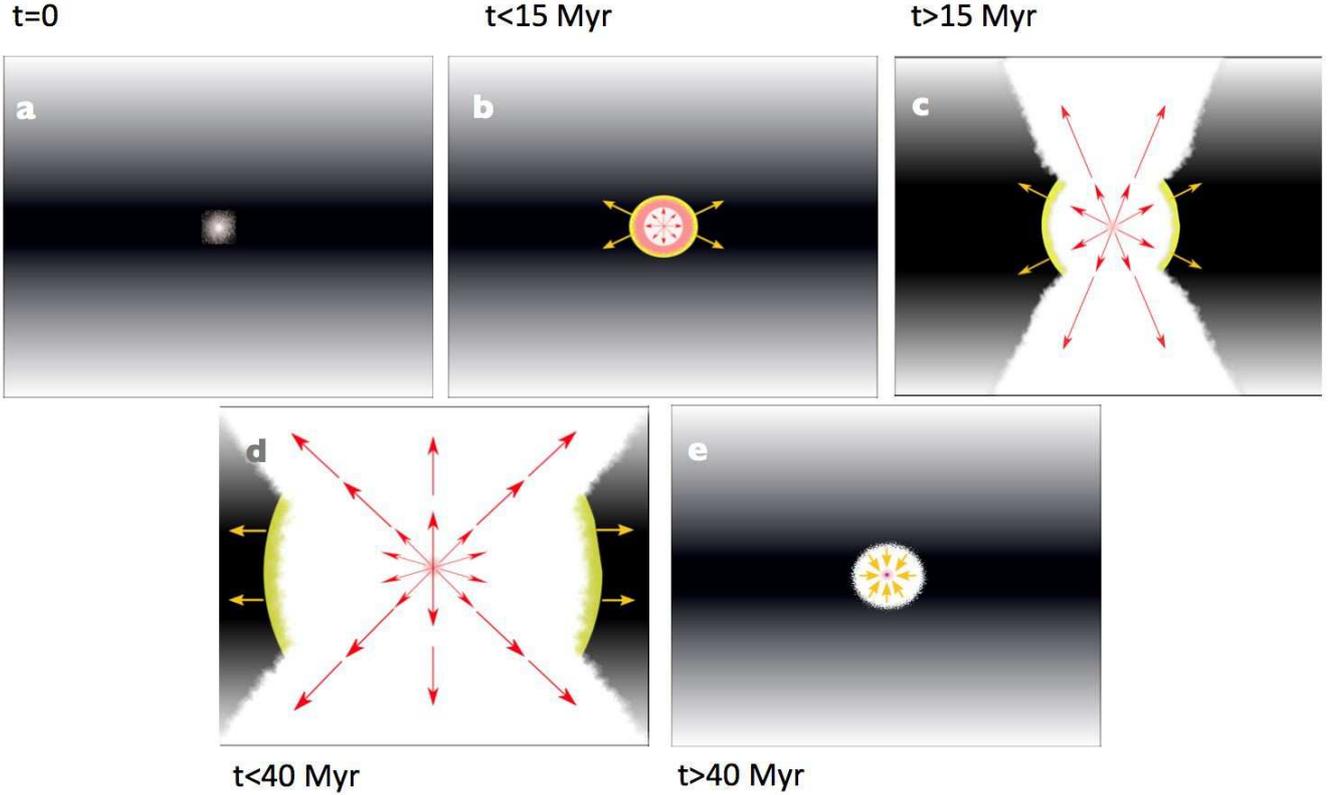}
}
%\vskip -15pt
\caption{{   This figure is a very schematic illustration of the evolution 
throughout the ``standard" formation of a two-generation cluster, described 
by the text equations (we point out  that the symmetry in the figure provides 
just a simplified view; the actual 3D configuration may be more complex 
because of the turbulence).}
    Times are indicative of the cluster age. Panel $a$: birth of the
    first stellar generation in the midplane of the stratified medium
    (grayscale) of a young galactic disk; panel $b$: beginning of the
    SN~II epoch, with expansion of the bubble in the disk. The local
    gas swept up by the bubble is confined in a cool, dense, thin
    shell (yellow), while the free expanding wind powered by the SNs
    (red arrows) is thermalized (red area) after crossing the reverse
    shock due to the reflection against the contact surface between
    the ambient gas and the SN ejecta; panel $c$: the bubble expands
    beyond the disk scale height and the SN wind is lost from the
    disk; panel $d$: the hole on the disk expand as long the SNs keep
    exploding. Depending on the parameters, it may still increase in
    size for a while after the SN activity, until its expansion
    velocity slows down to the velocity dispersion of the surrounding
    gas; panel $e$: after the end of the SN~II explosions, the disk
    matter closes up again into the cluster. 
    Depending on the bubble
    radius, there is time, or not, to form an extreme second
    generation from the super--AGB ejecta; then star formation
    includes AGB matter partially diluted with the re-accreted disk
    gas. Phases described in panels $b$ and $e$ may also represent the
    further stop of star formation due to binary delayed SN~II
    explosions (panel $b$) and the successive reaccretion phase, which
    now includes both the AGB and SN ejecta (panel $e$). } 
  
\label{fig:disegni} 
\end{figure*}

The model follows the steps illustrated in Figure \ref{fig:disegni}:
(a) the globular cluster is assumed to form within the disk of a
high-redshift galaxy \citep[not necessarily in its nucleus,
e.g.][]{kravtsov05, bekki12}; (b) the SNe II belonging to the FG start
to explode powering a wind which drives a bubble; (c) the bubble
eventually breaks out of the disk and the wind is almost all lost into
the space surrounding the galaxy; {   (d) the hole in the disk carved
  by the wind expands to its maximum radius $R_{\rm st}$, when its
  expansion velocity decreases down to the local velocity dispersion
  $\sigma_0$ of the surrounding gas; (e) at the end of the SN activity
  the ambient matter then penetrates the cavity in a time of the order
  of $R_{\rm st}/\sigma_0,$ and the disk gas, not affected by
  supernova contamination, dilutes the AGB ejecta out of which the SG
  Intermediate stars form.  }

In our simple model the disk is
described by a stratified gas distribution characterized by its
effective scale height
\begin{equation}
\he=\frac{1}{\rho_0}\int_0^\infty \rho(z){\rm d}z,
\end{equation}
where $\rho_0$ is the density in the midplane. The pressure includes
the turbulent pressure, and is defined as $P_0=\rho_0\sigma_0^2$; this
is a generalization of the thermal pressure because the total velocity
dispersion is defined as $\sigma_0^2=c_0^2+\upsilon_0^2$, where $c_0$
is the 1D thermal velocity dispersion and $\upsilon_0$ the 1D
turbulent velocity dispersion \citep[e.g.][]{lequeux05}.
\citet[][]{koyama09} have shown that the inclusion of the contribution
of the turbulence is necessary to support the common assumption that
the interstellar medium of the Milky Way disk is in effective
hydrostatic equilibrium. Moreover, the role of the turbulence prevails
in star bursting galaxies like those where the GCs presumably formed ,
where the velocity dispersion might be as high as $\sim 100$ km
s$^{-1}$ (but see section \ref{sec:disc}).

We assume that the globular cluster is located in the midplane, and
that its FG stars are characterized by a Kroupa \citep{kroupa01}
initial mass function (IMF); the minimum mass of the SN II progenitors
is M$_{\rm mass}$=8--9 $\msun$, while the most massive exploding stars
have $M=20$ $\msun$, as stars with higher masses may collapse quietly
to black holes \citep{smartt09}.  Assuming that the FG forms in an
instantaneous burst, the SN explosions are thus expected to last for a
period $t_{\rm SN}$ given by the lifetime difference between the
smallest and largest SN progenitor. Considered the uncertainty about the
minimum mass of such progenitors, and the details in stellar evolution
which weight on the value of M$_{\rm mass}$, we adopt $t_{\rm SN}\sim
25$ Myr as a reasonable compromise.

In general for massive objects like GCs a high explosion rate is
expected for the SNe belonging to the FG; as a consequence, most of
the SN energy input is thermalized, and a strong wind is driven out of
the cluster \citep[e.g.][]{clegg85}.  We, thus, consider bubbles
inflated by a continuous wind rather than by a sequel of instantaneous
explosions. On the numerical side, this treatment is supported by the
findings of \citet{maclow88} who showed that discrete explosions can
be approximated as continuous events when more than nearly seven SNe
explode one after the other. As far as the temporal wind evolution is
concerned, \citet{leitherer14} showed that SNe belonging to an
instantaneous burst release mass and energy at rates only slowly
decreasing in time; within our schematization we can safely assume
these rates to be constant, and we computed them as follows.

The total number of explosions is $N_{\rm SN}=7\times 10^{-3}(M_{\rm
  FG}/\msun)$, and the total amount of SN ejecta is $M_{\rm
  SN,ej}=0.09M_{\rm FG}$\footnote{Following \citet{iben84}, we assume
  that each SN leaves a neutron star of 1.5 $\msun$.}.  We thus obtain
$\dot M_{\rm SN}=M_{\rm SN,ej}/t_{\rm SN}$ and $L_{\rm SN}=N_{\rm SN}
E_0/t_{\rm SN}$, respectively, where $E_0=10^{51}$ erg is the energy
of a single SN. In summary, in our models we assume
\begin{equation}
\label{eq:snmdot}
\dot M_{\rm SN}=3.6\times 10^{-9}\left ({M_{\rm FG}\over \msun}\right )\; \msun\;{\rm yr^{-1}},
\end{equation}
\noindent
and
\begin{equation}
\label{eq:snedot}
L_{\rm SN}=9.0\times 10^{33}\left ({M_{\rm FG}\over \msun}\right )  \;{\rm erg\;s^{-1}}.
\end{equation}
\noindent
Assuming that all the SN luminosity is transformed in the wind
mechanical luminosity \footnote{Following \citet[cf.][]{calura15} we
  assume that the residual gas out of which the GC formed is expelled
  from the cluster by the SN wind.} $L_{\rm w}$, we obtain a wind
terminal velocity $V_{\rm w}=(2L_{\rm w}/\dot M_{\rm SN})^{1/2}=2800$
km s$^{-1}$.

\section{Interaction of the GC wind with the surrounding medium}
\label{sec:snrint}
\subsection{Bubble expansion: summary}
\label{subsec:bubexp}
Before describing the interaction of the GC wind with the surrounding
medium, we briefly summarize the evolution of a ``standard" spherical
bubble, that is, a bubble powered by a constant luminosity wind and
expanding through a local uniform gas \citep{weaver77}. The wind
generated by the sequential supernova explosions produces a hot bubble
which expands and sweeps out the ambient medium; this gas is
compressed into a shell by a shock located at its leading edge.  The
kinetic energy of the wind is partly converted into thermal energy by
a reverse shock located behind the shell, close to the centre. The
shocked ambient medium and the shocked hot wind are separated by a
contact discontinuity. {  In absence of relevant radiative losses, the
bubble expands in the energy-conserving regime, and its radius grows
in time as $R_{\rm b}\propto t^{3/5}$.  Given its high temperature and
low density, usually the bubble interior does not cools radiatively; 
instead, the outer shell, denser and cooler, is
radiative, and usually collapses quite soon. However, radiative
  losses from the shocked ambient medium have little effect on the
  expansion of the bubble which preserves the same time dependence,
  just reducing its expansion velocity of about 13\% \citep{weaver77}.
  For this reason in the following we still refer to this stage,
  although somewhat improperly, as to the ``energy-conserving
  regime".}

Under certain conditions, it may happen that the radiative cooling of
the bubble hot interior is large enough to affect the bubble
expansion. In this case, the thermal pressure of the shocked wind is
essentially absent, and the reverse shock moves close to the
shell. The wind energy in such a radiative bubble is not conserved,
but the wind momentum is: the wind pushes directly onto the shell,
transferring its momentum. This is the so-called momentum-conserving
regime, with $R_{\rm b}\propto t^{1/2}$ \citep{steigman75}.

Finally, we warn the reader that, to make computations
  easier, in the following the time $t$ is computed starting from the
  beginning of the SN explosions. Therefore, the age $t_{\rm age}$ of
  of the cluster is $t_{\rm age}+t_0$, where $t_0\sim 10$ Myr is the
  lifetime of a star with a mass of 20 M$_{\odot}$. 

We now go back to our model and see how the above concepts come into play.
\subsection{First expansion stage: spherical expansion}
\label{subsec:stg1}
The bubble preserves a nearly spherical shape as long as it expands
within the disk. During the expansion the pressure of the bubble
decreases, and two possibilities are in order: $i$) if the pressure
decreases down to the interstellar matter (ISM) pressure when the bubble radius is smaller
than $\he$ the bubble stalls. At this point, since the
  bubble is in a gravitational field, it is likely to be subject to the
  Rayleigh-Taylor instability; as a consequence, the ambient gas can mix with
  the bubble interior, leading to the cooling and collapse of the bubble itself; 
  $ii$) if the radius
reaches $\he$ the bubble breaks out.  Following \citet{koo92}, the
characteristic wind luminosity that determines whether a bubble will
be able to break out of the disc is
\begin{equation}
\label{eq:lcr}
L_{\rm br}=53.7\rho_0 \he^2\sigma_0^3\;\;{\rm erg\;\;s^{-1}},
\end{equation}
\noindent
where $\rho_0$ is the midplane density.
For the moment we consider cases for which $L_{\rm w}>L_{\rm br}$, 
leading to bipolar flows often observed in starburst galaxies \citep[e.g.][]{strickland09},
and postpone the discussion of the cases in which the opposite is true
(see sect. \ref{subsec:snmain}). This first stage ends at the time at
which the breakout occurs \footnote {Following \citet{maclow88},
it can be shown that, given the values of the parameters, usually the bubble is still
in the energy-conserving stage when it breaks out.} (see Appendix \ref{sec:app})
\begin{equation}
t_{\rm br,6}=0.19\hee^{5/3}\left ({L_{41} \over n_0}\right )^{-1/3},
\end{equation}
\noindent
where $n_0$ is the numerical density of the ambient gas,
$t_{\rm br,6}\equiv t_{\rm br}/{\rm Myr}$, 
$L_{41}\equiv L_{\rm w}/(10^{41} {\rm \;erg \; s^{-1})}$, and $\hee \equiv \he /10^2$ pc.
\subsection{Second expansion stage: wind momentum conservation}
\label{subsec:stg2}
At the breakout the shell suddenly accelerates. This acceleration
causes the disruption of the shell by the Rayleigh-Taylor
instability, and the wind blows directly out of the planar medium
\citep[e.g.][]{koo92}. As a consequence, the SN ejecta are quickly
lost by the galaxy, without any significant pollution of the local
ISM; this is a crucial point for the chemical evolution of the
cluster, as the different stellar populations within the majority of
the GCs do not show appreciable differences in the abundance of
iron. At the breakout the bubble pressure drops as the hot interior
leaks out. At this stage the wind impinges directly on the inner
surface of the portion of the shell which is moving parallel to the
plane; this shell, which we assume cylindrical, expands now in a wind
momentum-conserving regime.  

{ 
The shell radius grows to a maximum values,
the stalling radius $R_{\rm st}$, where the shell velocity
drops to the dispersion velocity $\sigma_0$ of the ambient medium; $R_{\rm st}$
is equal to $R_{\rm eq}$, the radius where the wind ram
pressure equals the pressure of the ISM. The expansion ceases and the
shell, whose density is now similar to that of the local gas, stalls
merging into the ISM.

It is possible, however, that the shell does not stalls at the
equilibrium radius.  This can occur because when the shell switches to
the wind momentum-conserving regime at a radius $\sim \he$, it has a
momentum given by the energy-conserving solution that is higher than
the momentum expected at that radius in a ``pure" momentum-conserving
regime (when the expansion is momentum conserving from the
beginning). Therefore the shell overruns the equilibrium radius
because its velocity is higher than $\sigma_0$, and keeps expanding
although the wind ram pressure is lower than the ambient pressure.
Later, when the shell velocity becomes equal to $\sigma_0$, the hole
stalls.

It may happen that the shell is still expanding at $t>t_{\rm SN}$,
when the wind has already ceased to blow.  In this case the hole
enters a third dynamical stage.  }

\subsection{Third expansion stage: shell momentum conservation}
\label{subsec:stg3}
This third stage starts at $t_{\rm SN}$, when the supernovae cease to
explode. Even in absence of the wind the shell continues to expand
because of the inertia of the swept-up mass. Eventually, however, it
stalls at $R_{\rm st}$.

\subsection{Analytical description of the expansion phase}
\label{subsec:ande}
In Appendix \ref{sec:app} we present a derivation of the the expansion
law of the bubble for each of the three dynamical stages:

\begin{equation}
 \rbb(t) = 
  \begin{cases} 
   2.7{\cal L}^{1/5}t_6^{3/5} & \text{if } t \leq t_{\rm br} \\
   \hee\left (t\over t_{\rm br} \right )^\beta  & \text{if } t_{\rm br} <  t \leq t_{\rm SN} \\
   (3\beta {R_{b,2}^3(t_{\rm SN,6})\over t_{\rm SN,6}}(t_6-t_{\rm SN,6}) +\\
   + R_{b,2}^3(t_{\rm SN,6}) )^{1/3}       & \text{if } t>t_{\rm SN}
  \end{cases}
\end{equation}

Here lenghts are in 0.1 kpc and times in Myr. The dependence of the bubble 
expansion on $L_{41}$ and $n_0$ occurs only through the ratio
${\cal L}\equiv L_{\rm 41}/n_0$, which is the main parameter
regulating the bubble evolution; this evolution depends also on a
second parameter, $\he$, which however plays a secondary role, as
shown in Figure \ref{fig:rv}. Finally, $\beta$ has a weak dependence on 
${\cal L}$ and $\he$, as described in the Appendix \ref{sec:app}.

As an example, we show in the lower panel of Fig. \ref{fig:rv} a horizontal dotted 
line corresponding to a dispersion velocity $\sigma_0=1.15\times 10^6$
cm s$^{-1}$ (equivalent to the isothermal sound speed of the ambient
medium whose temperature is assumed to be $10^4$ K). The intersection
of this line with the curves describing the expansion velocity of
different models determines the time $t_{\rm st}$ at which such models
reach the radius $R_{\rm st}$ and stop their growth.

\begin{figure}    
\centering{
 \includegraphics[width=8cm]{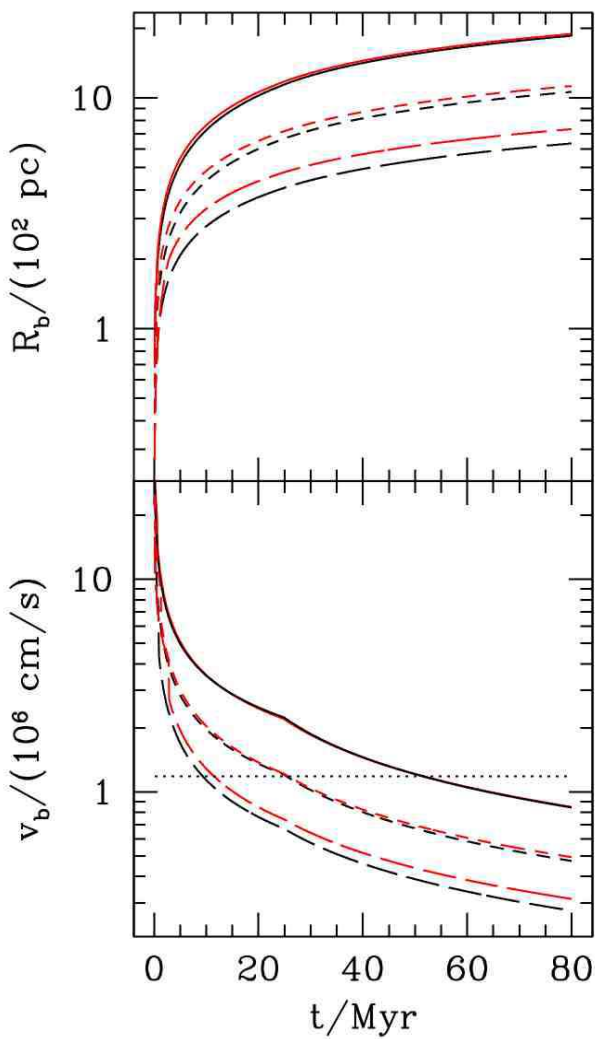}
}
\caption{Bubble radii (upper panel) and expansion velocities (lower
  panel) for different values of the parameters: ${\cal L}=0.01$
  (long-dashed lines), ${\cal L}=0.1$ (short-dashed lines), ${\cal
    L}=1$ (solid lines); black lines are for $\hee=1$, and red lines
  for $\hee=2$. The horizontal dotted line in the lower panel
  indicates the velocity dispersion of the ambient medium.  }
\label{fig:rv} 
\end{figure}

\subsection{Bubble contraction}
\label{subsec:bubcol}
{  
As pointed out in sect. \ref{subsec:stg2}, when the bubble reaches the radius $R_{\rm st}$
it loses its particular structure, and can be simply regarded as a
hole in the disk rather than a bubble; however, in order to avoid
confusion with symbols and nomenclature we continue to refer to such
hole as ``bubble". At $R_{\rm st}$ the wind ram pressure is inadequate
to hold the bubble at this radius, and the ambient gas encroaches into the
cavity with velocity $\sigma_0$ \citep{heiles87,{dercole92},koo92}; in the
following we will refer to this stage of progressive replenishment and reduction of the
hole as ``bubble contraction'' or ``bubble shrinking".
}
 
If the GC wind is still active (that is, if $t<t_{\rm SN}$), a steady
state configuration is established at the equilibrium radius where the
ram pressure of the wind balances the ambient pressure, $\rho_{\rm
  w}(R_{\rm eq}) V_{\rm w}^2=P_0$:
\begin{equation}
\label{eq:eqr}
R_{\rm eq,2}=41.43\left ({{\cal L}\over V_{\rm w,8}\sigma_{0,6}^2}\right )^{1/2},
\end{equation}
\noindent
where $V_{\rm w,8}\equiv V_{\rm w}/10^8 \; {\rm cm\;s^{-1}}$, and $\sigma_{0,6}\equiv \sigma_{\rm 0}/10^6\; {\rm cm\;s^{-1}}$.
Once the GC wind ceases, at $t\geq t_{\rm SN}$, the bubble shrinks 
again completing its contraction at the time 
\begin{equation}
\label{eq:tco1}
t_{\rm co}=t_{\rm SN}+R_{\rm eq}/\sigma_0.
\end{equation}
For higher values of ${\cal L}$, the receding bubble reaches $R_{\rm
  eq}$ at $t>t_{\rm SN}$, and therefore contracts directly toward the
centre after
\begin{equation}
\label{eq:tco2}
t_{\rm co}=t_{\rm st}+R_{\rm st}/\sigma_0.
\end{equation}

\begin{figure}    
\centering{
 \includegraphics[width=8cm]{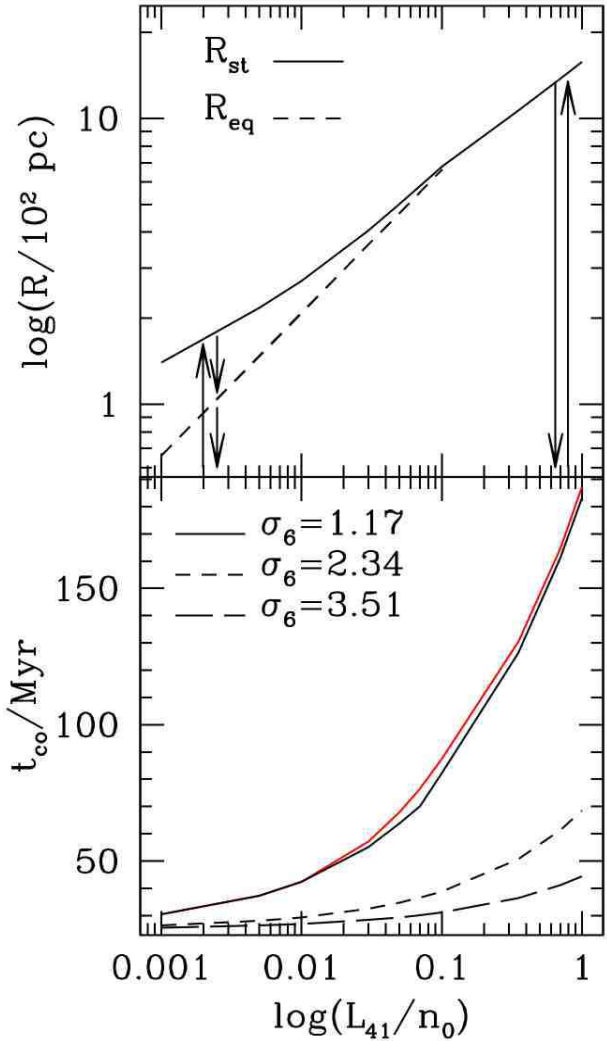}
}
\caption{Upper panel: stalling radius $R_{\rm st}$ (solid line) and
  equilibrium radius $R_{\rm eq}$ (dashed line) as function of ${\cal
    L}$ for $\hee=1$. The behaviour of these radii for $\hee=2$ is
  similar, but we do not show them to avoid to clutter the figure. For
  the same reason we show only the case $\sigma_6=1.17$.  The arrows
  on the left side indicate schematically the behaviour of the radius
  of bubbles with ${\cal L }\leq 0.1$: it grows up to $R_{\rm st}$ and
  then shrinks reaching $R_{\rm eq}$ at a time $t<t_{\rm SN}$.  Here
  the bubble settles until $t_{\rm SN}$; later on it contracts
  definitively. The evolution of bubbles with ${\cal L}>0.1$ is
  schematized by the two arrows on the right side: in this case the
  bubble, after reaching its maximum size, contracts directly until it
  closes. For these bubbles, the portion of the dashed line extending beyond ${\cal
    L}=0.1$  is meaningless and is not shown in the figure (see text).    
  Lower panel: lifetime of a bubble as function of ${\cal L}$ for
  three values of $\sigma_0$. For the case $\sigma_0=11.7$ km s$^{-1}$
  we show the case $\he=100$ pc (black line) and $\he=200$ pc (red
  line).  }
\label{fig:rt} 
\end{figure}

The dynamical behaviour of the bubbles is summarized in
Fig. \ref{fig:rt}. As an example, we report in the upper panel just
the case with $\sigma_{0,6}=1.17$ and $\hee=1$; the general behaviour
of other cases is similar, but we do not show them in order to avoid
to clutter the figure. The solid line indicates the maximum radius
reached by bubbles with different ${\cal L}$, while the dashed line
shows the equilibrium radius at which bubbles with ${\cal L}\leq 0.1$
settles during the successive contraction, ``waiting" for the end of
the SN activity; after $t_{\rm SN}$ the contraction is restored until its
conclusion. This expansion and the successive ``two--step"
contraction are represented by the three arrows in the low ${\cal L}$
region of the diagram. The two arrows on the right side show that
bubbles with ${\cal L}>0.1$ contract totally and directly after
reaching their maximum extension. For these bubbles the dashed line
indicating the equilibrium radius is meaningless; we show only the
portion of this line for ${\cal L}\leq 0.1$ to point out that only
this segment is meaningful. The critical value ${\cal L}=0.1$
separating the behaviour of the collapsing bubbles varies for
different values of the parameters $\sigma_0$ and $\he$, but the overall
bubble dynamics remains qualitatively unchanged.

The lower panel of Fig. \ref{fig:rt} shows the lifetime of bubbles
with different ${\cal L}$. It has been computed by equation
(\ref{eq:tco1}) or equation (\ref{eq:tco2}), depending on the value of
${\cal L}$.

\section{Discussion}
\label{sec:disc}
Figure \ref{fig:rt}, together with the results of previous simple
chemical evolution models based on the AGB scenario
\citep{der10,der12}, constrain  the ambient medium properties needed 
 to have a viable dilution mechanism. As shown in \citep{der10,der12}, 
 in order to form a suitable amount of E stars --
for example, 18\% of the present GC mass as in NGC 2808
\citep{carretta09} -- pure undiluted AGB ejecta must flow in
the cluster centre for about 20 Myr after the end of the SN II
activity.  After this time a substantial episode of dilution is needed
to form the Intermediate population; in NGC 2808 it represents about
32\% of the total, and nearly 40--50 \% of its mass is composed by
accreted pristine gas.

The above requisites indicate that the bubble must have a lifetime of
about 45 -- 50 Myr; from Fig. \ref{fig:rt} we see that, for the model
with $\sigma_{0,6}=1.17$, it must be $0.01 \simlt {\cal L} \simlt 0.02$,
and that the maximum extension reached by the bubble for $\hee=1$
($\hee=2$) is $R_{\rm st}\sim 300$ pc ($R_{\rm st}\sim 400$ pc).
Assuming $M_{\rm FG}=10^7$ M$_{\odot}$, it turns out that $\rho_0\sim
1.5-2.0$ $\msun$ pc$^{-3}$, $\Sigma\sim 300-800$ $\msun$ pc$^{-2}$,
and $P_{0,6}\sim 1$, where $P_{0.6}\equiv P_0/(10^6\,\kappa\;\,{\rm
  cm^{-3}\,K)}$, and $\kappa$ is the Boltzmann constant.

The above figures are order of magnitudes larger than those of the
Milky Way. Yet, they are at the low end of gas-rich, normal galaxies
at $z>2$ which may reach pressures $P/\kappa \simgt 10^8$ K
cm$^{-3}$, $\sigma_0=10-100$ km s$^{-1}$ (but see sect. \ref{subsec:snmain}), and surface
densities $\Sigma > 10^3$ $\msun$ pc$^{-2}$ \citep[cf.][and
references therein]{kruijssen15}. 

Our result shows that our model requires properties of the ambient
surrounding GCs during their formation and early evolution consistent
with those of ``normal", star-forming disk galaxies observed at
redshift $z>2$, a conclusion in agreement with the findings of
\citet{kruijssen15}.

We have discussed in detail the model with $\sigma_{0,6}=1.17$
illustrated in the lower panel of Figure \ref{fig:rt}.  In the same
panel we also show the, possibly more realistic, cases with
$\sigma_{0,6}=2.34$ and $\sigma_{0,6}=3.51$. Increasing $\sigma_0$
leads to values of ${\cal L}$ appropriate for a NGC 2808-like GC which
span a broader range and are shifted to higher values.

Models with low values of ${\cal L}$ are suitable to describe GCs
where the dilution is expected to occur quite soon (i.e. clusters in
which the Extreme population is absent, as M4 (NGC 6121)
\citep{der10}.  The sequence of events for these clusters would be
identical to those exhibiting an extended Na--O anticorrelation
(cf. Fig.~\ref{fig:disegni}) with the only difference being the
smaller size reached by the bubble.

Models with large values of ${\cal L}$ require a long time to shrink
(see Fig. \ref{fig:rt}). If this time is longer than the time after
which delayed SNe~II or Type~Ia supernovae (SNe~Ia) start to explode
(see sect. \ref{subsec:delsn}), the bubble does not contract
completely and the cluster does not accrete pristine gas. In this case
the GC should host a SG population composed only by extreme stars
formed out of the undiluted AGB ejecta during the time interval
between the end of the single SNe~II activity and the beginning of the
late SN~II and/or SN~Ia explosions. \citet{dantona16} hypothesize that
this could be the case of the globular cluster NGC 2419 for which the
results of \citet{dicriscienzo15} and \citet{cohen11} suggest that the
second generation is entirely or mostly Extreme.

More in general, being the bubble lifetime an inceasing function
  of ${\cal L}$, and thus of the initial cluster mass, a correlation
  occurs between the strength of the signature of the AGB chemical
  pollution and the cluster mass.  This correlation may be weak, as we
  have seen that this lifetime depends also on several other physical
  inputs, but nevertheless it should be present. A particular case
  concerns the helium content of the second generation stars.  We
  noticed above that the presence of a high helium (Y$\sim$0.35-0.38)
  population in the most massive clusters is the outcome of stars from
  undiluted AGB ejecta. In addition, as a consequence of the model
  presented here we expect a correlation between the {\it average}
  helium content of the second generation and the cluster mass. In
  fact, clusters with smaller masses are characterized by smaller AGB
  ejecta rate and shorter reaccretion time, and fewer or no stars may
  form from gas in which the ejecta are not strongly diluted.

Other evolutionary paths for the bubble are also possible. In fact,
from equation (\ref{eq:lcr}) we know that, if ${\cal L}\simlt {\cal L}_{\rm
  br}\equiv 5\times 10^{-5}\hee^2 \sigma_{0,6}^3$, the bubble is not
able to break-out and its evolution is rather different from that
described in section \ref{sec:mod}; is this a realistic
option? In the following, we discuss two main cases.

\subsection{The SN II main phase}
\label{subsec:snmain}
Given its strong dependence on $\sigma_0$, high values of ${\cal L}_{\rm
  br}$ are attained for large values of the velocity dispersion which
may be as high as $\sigma_{0,6}=10$ in star forming galaxies at $z=2$.
This value, however, is rather extreme; \citet{wisnioski15} shows that
the mean velocity dispersion is around $\sigma_{0,6}=5$, and a similar
result is obtained by \citet{tacconi13}. Moreover, a possible lack of
an adequate resolution may lead to overestimated values of the
velocity dispersion \citep{newman13}.

However, despite the above arguments, and although GCs usually form in
``normal" galaxies, it cannot be excluded that in some clusters the
wind is not strong enough to satisfy the conditions required for a
break-out.  In this case the bubble, which remains approximately
spherical, expands up to the stalling radius, and keeps such size
until $t_{\rm SN}$, after that contracts in a time $\sim R_{\rm
  st}/\sigma_0$. The SN II ejecta are all contained within the bubble,
but they will not necessarily contaminate the gas out of which the SG
will form.

In fact, it is possible that the cluster is not exactly at rest
with respect to its ambient gas; even a low relative velocity is
enough to allow the cluster to move away from the centre of the bubble
which instead remains in its original position. It is also possible
that the cluster reaches the shell; the bubble then becomes distorted
and its leading edge acquires the shape of a bow-shock
\citep{weaver77,brighenti94}. In any case, at the end of the SN II
activity, the bubble contracts toward the centre, driving there its
content of SN ejecta; being the GC off-centre, the SN ejecta are not
involved in the SG star formation occurring in the cluster core
(see also sect. \ref{subsec:gcmov}).

\subsection{The delayed binary SN II phase}
\label{subsec:delsn}
It is well known that SN II may occur also after the end of the main
phase due to the evolution of single stars. In fact, in interacting
binaries having mass components below the minimum mass for explosion,
mass transfer may push the accreting component beyond such minimum
\citep{vankerkwijk99,portegies99,tauris00}.  The frequency of such
events may vary strongly from a cluster to another \footnote{Computing
  the frequency of delayed SNe~II and the extent of the time interval
  during which they explode requires specific stellar population
  synthesis models taking into account the fraction of primordial
  binary systems and their initial properties (e.g. period and mass
  ratio distribution).  More details on the role of the delayed SNe II
  within the AGB scenario can be found in \citet{dantona16}.}. 
  We do not consider the case in which ${\cal
  L_{\rm del SN~II}}\simgt {\cal L}_{\rm br}$, as it is evident that
this would fully preclude re-accretion, and also hamper the formation
of second generation stars for a long hiatus. We are here more
interested in the cases ${\cal L_{\rm del SN~II}}\simlt {\cal L}_{\rm
  br}$. We can distinguish different possibilities:

1) Delayed SN~II explosions isolated in time: as already discussed in
\cite{der08} one isolated SN explosion will not produce significant
dynamical effects: the small bubble will close soon enough and the
supernova ejecta will take part in the cooling flow and contribute to
the formation of new stars in the core. Anyway, one such SN~II
explosion will only disperse into the medium $\sim 0.07$ $\msun$ of
iron and a more relevant 0.5 $\msun$\ of oxygen. Whether or not this
material may cause a visible spread in the abundances of second
generation stars depends on the amount of polluted material and on the
initial cluster metallicity (the lower the metal abundances, the
larger will be the effect of oxygen pollution).

2) A phase of delayed SN~II explosions extended in time, lasting for
several 10$^7$yr: this case study may lead to the formation of
`s-Fe-anomalous clusters'.  While most GCs can be considered
mono-metallic, with an upper limit to the scatter of iron less than
0.05\,dex \citep{carretta2009c}, in recent years it has been found
that there is a small number of objects characterized by an intrinsic
internal [Fe/H] abundance dispersions \citep[see][for a summary
review]{dacosta2015}. The fraction of stars with increased iron
content may vary from a few percent to 15--30\% \citep{marino2015},
and the most intriguing characteristic is that the stars with larger
[Fe/H] abundances also have higher [s-element/Fe] abundance ratios,
requiring an additional s--process element contribution, most probably
from AGB--stars, over and above the contribution needed to maintain
the abundance ratio as the iron abundance increases
\citep{dacosta2015}. These clusters have been dubbed `s-Fe-anomalous'
by \cite{marino2015}.  {\it Both iron groups of `s-Fe-anomalous
  clusters' show independent O--Na anti correlation}.  We discuss in
\citet{dantona16} how these groups may be formed from a
chemical point of view. Here we show that our model provides a
dynamical scenario able to produce the observed chemical properties.

Let us assume that the cluster had time to form a `first' second
generation, after reaccretion of the disk gas (this is necessary to
produce the O--Na anticorrelation shown by the low metallicity stars),
and that, afterwards, at about 60--70Myr of age, the delayed SN~II
epoch halts the second--generation star formation for a time $t_{\rm
  delSN}$. The energy of each supernova will be, as for single
supernovae explosions, $E_0=10^{51}$ erg and we can assume that the
whole luminosity goes into the wind mechanical luminosity. We get
$L_{\rm w(del.SN)}=N_{\rm delSN}E_0/t_{\rm delSN}$.  The AGB ejecta
become s-process rich when their envelopes experience the effects of
numerous episodes of third dredge up, at $\sim$90--100 Myr of age. Thus
we have a typical timescale of $t_{\rm delSN}$=20--40~Myr. For a total
number of 10-40 explosions, we obtain $L_{\rm w,delSN} \simeq 0.8-6
\times 10^{37}$erg s$^{-1}$. This value is small enough that indeed
${\cal L}<{\cal L}_{\rm br}$, and the wind will not freely escape from
the disk. The bubble radius may even be so small that it will not
reach the cluster periphery \citep{der08}, but this will be sufficient
to prevent star formation, until the bubble recollapses into the core.
The matter forming this second phase of second generation formation
will be enriched both by the ejecta of the supernovae and from the AGB
winds, which will now be s--enriched and CNO enriched.

We can summarize the time sequence of the events leading to the
formation of s-Fe-anomalous clusters as follows:
\begin{enumerate}
\item Formation of SG stars with the iron content equal to that of the
  FG population: this follows the steps outlined in the paper and
  schematically illustrated in Figure \ref{fig:disegni}.
\item Delayed SN~II epoch: a bubble forms (panel b of Fig. 1) and
  remains confined into the disk, and possibly within the cluster
  periphery. The iron and oxygen of these supernovae remain confined
  and mix with the disk gas at the bubble periphery as a result of a
  number of possible processes, such that Rayleigh-Taylor instability
  (see sect. \ref{subsec:stg1}), turbulent mixing layers
  \citep{slavin93}, thermal evaporation \citep{weaver77} and
  photoevaporation\citep{mckee84}. The average metallicity of the gas
  surrounding the bubble increases to [Fe/H]$_0+ \Delta$[Fe/H].  Also
  the AGB winds coming from inside the bubble are locked with the disk
  gas. Star formation is off for 20--40Myr.
\item Recollapse of the bubble and second phase of second generation
  formation. A burst of star formation occurs from the gas polluted by
  the delayed SN~II and mixed with AGB winds. If the mixing is
  inhomogeneous, we may still obtain an O--Na anti correlation in
  stars having iron content [Fe/H]$_0+ \Delta$[Fe/H].
\item The timing of this event is such that it occurs close to the
  beginning of the SN~Ia explosions in the cluster, which ends the
  second star formation epoch.
\end{enumerate}

\section{Some more complexities}
\label{sec:compl}
It is important, in our model, that the disk gas surrounding the
cluster remains uncontaminated despite the metals ejected by the SNe;
in this way it can give rise, once moved within the cluster, to a
SG population with the same iron abundance of the FG one, as observed
in most of the clusters \citep{piotto15}. In this section we consider
some aspects of the model which have been neglected until now, and
which could lead to an increment of [Fe/H] of the disk gas. We show
that this gas effectively preserves its original chemical composition,
at least during the time intervals of interest for the model.

\subsection{Asymmetric models}
\label{subsec:asym}
Until now we considered symmetric models where the cluster is located
on the galactic midplane. Of course, the cluster may also form
off-plane, at a distance $0\leq h\simlt H$ from the middle of the
disk. In this case the associated asymmetric bubble would
preferentially break out of one side of the disk, but not necessarily
out of the other; a non negligible fraction of the SN ejecta would
then remain trapped within the ISM that, after accretion, would lead
to a SG enriched in Fe, at odd with the majority of the observed
galactic clusters. Of course, a critical height $h_{\rm c}$ exists,
above which the bubble is effectively unable to cross entirely the
disk thickness. This critical height depends, of course, on $L_{\rm
  w}$, $n_0$ and $H$, but finding out such dependence requires
hydrodynamic simulations which are beyond the scope of this
paper. However, we can get some insight into the behaviour of an
asymmetric bubble from the numerical models of \citet{maclow88}. These
authors considered a plane parallel exponential atmosphere with
$\hee=1$ and $n_0=1$ cm$^{-3}$ crossed by a bubble with ${\cal
  L}=10^{-3}$; they find $h_{\rm c}=0.6\he$, while for $h<h_{\rm c}$
the bubble blows out of the opposite side at $t<1.2$ Myr. Being ${\cal
  L}\sim 0.01$ in our models, we believe that the bubble easily erupts
from the distant side of the plane; scaling the Mac Low \& McCray
results, this would occur after about 0.6 Myr.

We also stress that a GC initially placed above the galactic plane is
expected to oscillate up and down across the disk with a period
\citep[e.g.][]{shaviv06}
\begin{equation}
\label{eq:osc}
P_0\simeq\left ({\pi\over G\rho_0}\right )^{0.5}=26.42\left ({\rho_0\over M_{\odot}\;{\rm pc}^{-3}}\right )^{-0.5}\;\;\;{\rm Myr}.
\end{equation}
\noindent
Thus, for a typical value $\rho_0=1$ M$_{\odot}$ pc$^{-3}$ the cluster
crosses the plane two times during the SN activity. These oscillations
represent a further degree of flexibility of the model which helps to
account for differences among the clusters.

\subsection{Potential pollution by the GC wind}
\label{subsec:polwin}
In our model we assumed that the ISM on the disk surrounding the
cluster is not enriched by the SN metals. We now discuss this
assumption. Numerical simulations
\citep[e.g.][]{schiano85,tenorio97,tenorio98} show that the wind
hitting the shell on the disk is deflected away from the plane,
ablating the superficial gas layers of the inner side of the shell and
dragging them away. \citet{maclow99} proved that the fraction of SN
metals retained by the ISM decreases with the increasing wind power,
and is substantially absent for $L_{41}>10^{-3}$.

In any case, the possible mixing achieved in numerical simulations
represents an upper limit because it depends on the numerical
diffusion which is usually much greater than the physical one
\citep[e.g.][]{marcolini04}. The physical diffusion is a rather slow
process in warm and cold gas, and this led \citet{tenorio96} to put
forward a scenario for the ISM enrichment in which the metal-enriched
blown-out gas remains in the hot rarefied gaseous halo where mixes
efficiently. This hot gas than cools, condenses and falls back onto
most of the disk evenly after about $10^8$ yr. Contrary
to other proposed mechanisms such as galactic fountains, turbulence
and accretion \citep[cf.][and references therein]{tolstoy09,melioli15}
this framework can explain the absence of significant metal abundance
gradients throughout the dwarf galaxies, and represents our main
current understanding of the ISM pollution.

We thus conclude that the strong winds considered in our model do not
significantly contaminate the disk, in particular for the interval of
time of interest, which is shorter than $10^8$ yr.

\subsection{Triggered star formation}
\label{subsec:trigsf}
The expanding shell may trigger star formation (SF) by gravitational
collapse of swept-up gas as well as by compression of pre-existing gas
clouds \citep[e.g.][]{elmegreen11}. The SNs belonging to this
triggered stellar population could contaminate the surrounding disk
gas with their iron; then this gas, once moved into the cluster at
the end of the GC wind, would contribute to the formation of a stellar
SG with a [Fe/H] higher than that of the FG.  Actually, this could
occur if all the iron released by SNs were ``immediately" available
for ambient gas pollution; this is not the case, however. Stars formed
within clouds engulfed and overrun by the expanding shell, as well as
stars formed within the dense shell, are exposed to the GC wind, and
the ejecta of the SNs belonging to this stellar population are quickly
vented away. Bubbles powered by clusters with ${\cal L}\simlt 0.1$
(see Fig. \ref{fig:rt}) stall (interrupting any star formation) when
the GC wind is still blowing, and no ISM pollution is expected. For
${\cal L}\simgt 0.1$ the shell keeps expanding for a while after the
SN stage; in this phase, however, the shell is only weakly supersonic,
and its effectiveness in triggering SF is most likely reduced
\citep{elmegreen11}. Moreover, part of the iron released by SNs formed
in this period should be presumably lost above the disk, at least for
a while (see sect. \ref{subsec:polwin}).  Finally, only a few clusters
are expected to have ${\cal L} > 0.1$; we can speculate that pollution
from these supernovae may have played a role in those clusters where
large iron spreads are found, such as $\omega$-Cen, M54 and possibly
NGC 6273 \citep[][and references therein]{yong16}.

{ 
\subsection{GC motion}
\label{subsec:gcmov}
So far we considered a cluster at rest along the disk relative to the
surrounding gas. This is a quite reasonable assumption; in fact, dwarf
galaxies rotate as solid bodies, allowing the GC to remain very near
its birth place on time-scales of $\sim 10^8$ yr \citep[][and
references therein]{Warren11}, shorter than those of interest in our
model. In any case, we discuss here some general aspects of the
influence of the GC motion. Being the GC wind supersonic, a bow
shock forms head of the cluster; the stagnation point, that is the
point where the ram pressure of the wind balances the ISM pressure
is located at the stand-off distance \citep[e.g.][]{draine11}
\begin{equation}
\label{eq:offd}
l_{\rm off,2}=41.43\left ({{\cal L}\over V_{\rm w,8}\left (\sigma_{0,6}^2+v_{\rm GC,6}^2\right )}\right )^{1/2},
\end{equation}
\noindent
where $l_{\rm off,2}$ is the stand-off distance in unit of 100 pc, and
$v_{\rm GC,6}$ is the GC velocity relative to tha ambient gas in unit
$10^6$ cm s$^{-1}$ (note that the above equation is a generalization of
  equation (\ref{eq:eqr}) taking into account the ISM ram
  pressure). Once the wind stops blowing, the surrounding gas reaches
  the GC in a time $l_{\rm off}/(v_{\rm GC}+\sigma_0)$, and
  successively overruns it. As the ISM flows through the cluster, part
  of it can be accreted provided that $\sigma_*^2\simgt
  \sigma_0^2+v_{\rm GC}^2$, where $\sigma_*$ is the central stellar
  velocity dispersion of the cluster \citep[e.g.][]{naiman11}.  We
  conclude that the delayed accretion may occur earlier, but is not
  hampered by the motion of the cluster.
  
  {   To give an even approximate estimate of the mass accreted by
    the cluster is not an easy task. While for a point mass accretor
    the mass accretion rate can be assessed by the
    Bondi-Hoyle-Lyttleton formula \citep{bondi52}, an analogous simple
    analytical formula for the accretion onto core potentials does not
    exist, and one must resort to hydrodynamic
    simulations. \citet{naiman11} performed a number of 2D numerical
    simulations appropriate for different astrophysical contexts. They
    find that, in general, the accumulation of external gas occurs
    during an initial period lasting 10-100 sound crossing times
    $t_{\rm cr}$ of the cluster; eventually a steady configuration is
    established, and the gas accumulated into the cluster ceases to
    grow appreciably \footnote{{   The isothermal models of
        \citet{naiman11} are obtained assuming a specific heats ratio
        $\gamma=1.01$. If a full treatment of the radiative cooling
        were considered, then a steady state may not exist (Naiman,
        private communication).}}.
  
The parameters of model A of \citet{naiman11} resemble those met in
our paper ($v_{\rm GC}=2\sigma_0=2\times 10^6$ cm s$^{-1}$, $n_0=100$
cm$^{-3}$) although $M_{\rm GC}=3.5\times 10^5$ M$_{\odot}$ is lower
than the cluster masses considered in the present article. It turns
out that the accreted mass is $M_{\rm acc}=0.05M_{\rm GC}$, and that
this accretion occurs in a time $\sim 10-100\; t_{\rm cr}\sim 1-10$
Myr. Following
\citet{naiman11}, the gas accumulated into the core may reach
densities as high as $10^4\rho_0$; the local free-fall time is then
$\sim 0.035$ Myr, and stars are expected to form rather quickly.

Although this is  a very promising result as it suggests that the cluster can 
indeed re-accrete  the  correct  amount of needed pristine gas 
\citep[see, e.g.,][]{der10, der11,der12}, we strongly emphasize that 
additional simulations are necessary which include the 
 gas shed by the AGB stars as
well as the formation of SG stars, two ingredients which are absent in
the \citet{naiman11} models\footnote{{   Simulations taking into
    account sources and sinks of gas can be found in literature, but
    they are intended to study the accretion of intracluster medium
    onto galaxies in clusters \citep[e.g.][and references
    therein]{balsara94}, thus exploring a region of the parameter
    space of no interest here.}}.

}
\subsection{Multiple GCs}
\label{subsec:gcmul}
The model presented in the previous sections refers to the evolution
of a single bubble.  Although the formation of isolated GCs cannot be
excluded \citep{bekki12}, the formation of numerous clusters can
effectively occur. We have seen that the typical time-scale for the
evolution of a single bubble is $\tau\simlt 10^8$ yr; after this time
the ISM essentially recovers its initial distribution on the
disk. Therefore, if a second cluster forms after a time longer than
$\tau$, the two bubbles evolve individually.  If, instead, more
clusters form at the same time, the single bubbles coalesce forming a
superbubble which eventually contracts at the end of the SN stage. As
the superbubble shrinks, the clusters wthin the hole are overrun by
the flowing ISM, accreting some of it as discussed in
sect. \ref{subsec:gcmov}. Also in this case a delayed accretion
occurs.

}

\section{Conclusion}
\label{sec:conc}
We proposed a scheme for the interaction of the GC wind powered by the
SNe II with the surrounding ambient medium. This interaction provides
a viable mechanism to supply pristine gas and dilutes AGB ejecta
during the SG star formation phase in early Globular Clusters.

We have shown that our proposed mechanism naturally fulfills all the
requirements necessary to produce the observed Na-O anticorrelation in
the context of the AGB model. Initially, the spherical SN wind drives
the expansion of a bubble that, in most cases, breaks out of the
gaseous disk where the GC is located. As a consequence, almost all of
the SN ejecta is lost along the vertical direction, while the portion
of the wind moving mainly parallel to the midplane continues to drive
the growth of a hole through the gaseous disk. Some time after the SN
activity, the hole starts to contract carrying the surrounding ambient
gas, {\it with pristine abundances}, toward the cluster. The contraction
takes a certain period of time during which the ejecta of the massive
AGB stars collect within the GC centre and Extreme, O-poor, He-rich
stars can form. Once the ambient medium reaches the cluster, it mixes
with the AGB ejecta, and Intermediate stars form out of this mixture.
 
The results of the mechanism presented here are consistent with some
still unexplained observational results. In particular, in agreement
with observations, only clusters with large ${\cal L} = L_{\rm
  w}/n_0$, and thus mainly clusters with large initial masses, can
form an extended Na-O anticorrelation, while the lower values of
${\cal L}$ expected for clusters with lower masses do not allow the
formation of O-poor, He-rich Extreme stars.  In general, a
  correlation between the spread in helium and the cluster mass is an
  outcome of the proposed model.  Moreover, the pristine gas involved
in the dilution is not enriched in iron, giving rise to a SG with the
same [Fe/H] of the FG stars, as often found in the observed GCs.

We finally conjectured that the presented scenario may also explain
the s-Fe-anomalous clusters where the stars are divided into two main
groups with different iron abundances, each group exhibiting an
independent O--Na anticorrelation, and with the iron--richer group
showing an enhancement of s--process elements. In fact, assuming that
few tens of delayed SN II start to explode a short while after the end
of the single SNe II explosions, and supposing that such activity
lasts for 20--40 Myr, a second bubble forms which, however, stalls
before breaking out. The SN ejecta remain trapped into the bubble, and
fall back into the cluster when the bubble contracts. This leads to
the formation of Intermediate stars in the same way as the collapse of
the first bubble did; but the matter forming this second burst of SG
stars is enriched with the SN ejecta, giving rise to a bimodality in
the iron distribution of the stellar populations.

\section*{Acknowledgements}
We thank the anonymous referee for her/his useful comments.
Fruitful discussions with F. Fraternali, J.P. Naiman, R. Sancisi, M. Tosi and G. Zamorani are 
also acknowledged. F. D'A. and A. D'E. acknowledge support from PRIN INAF    
2014 (principal investigator S. Cassisi). E.V. acknowledges support from 
grant NASA-NNX13AF45G.

%%%%%%%%%%%%%%%%%%%%%%%%%%%%%%%%%%%%%%%%%%%%%%%%%%



\bibliographystyle{mnras}
\bibliography{arx_ref}

%%%%%%%%%%%%%%%%%%%%%%%%%%%%%%%%%%%%%%%%%%%%%%%%%%

%%%%%%%%%%%%%%%%% APPENDICES %%%%%%%%%%%%%%%%%%%%%

\appendix

\section{Bubble expansion in a stratified medium} 
\label{sec:app}

Here we consider the interaction of a GC wind with a surrounding
stratified medium with an effective scale height $\he$ and a midplane
density $\rho_0$. The GC wind drives a bubble whose expansion can be
divided in three stages.

As long as the bubble radius is shorter than $\he$, it can be assumed
that the bubble is expanding in an uniform medium with density
$\rho_0$, and its radius $R_{\rm b}$ is given by \citep{weaver77}
\begin{equation}
\label{eq:rb1}
R_{\rm b,2}(t)=2.7\left ({L_{41} \over n_0}\right )^{1/5}t_6^{3/5},
\end{equation}
\noindent
where $R_{\rm b,2} \equiv R_{\rm b}/\,100 \,{\rm pc}$, $L_{41}$ is the
wind luminosity in units of $10^{41}$ erg s$^{-1}$, $n_0$ is the
numerical density of the ambient medium in cm$^{-3}$, and $t_6$ is the
time in Myr.  As the wind is very powerful, the bubble eventually
breaks out at the breaking time
\begin{equation}
t_{\rm br,6}=0.19\hee^{5/3}\left ({L_{41} \over n_0}\right )^{-1/3}
\end{equation}
\noindent
obtained from equation (\ref{eq:rb1}) for $R_{\rm b}=\he$.

This time marks the end of the first stage and the beginning of the
second dynamical phase. At the breakout the shell suddenly
accelerates. This acceleration provokes the disruption of the shell by
the Rayleigh-Taylor instability, and the wind blows directly out of
the planar medium \citep[e.g.][]{koo92}. From this time forth we focus
on the ``cylindrical hole" in the planar gas which still keeps
expanding under the action of the fraction $f(R{\rm b},\he)$ of the
isotropic wind intercepted by its internal wall.  After the breakout,
the hot internal gas leaks out and the internal pressure of the bubble
drops; the wind thus impinges directly on the cylindrical shell which
thus expands now in a momentum-conserving regime. During this second
stage, the equation regulating the motion of the shell is
\begin{equation}
\label{eq:eqrb2}
2\pi R_{\rm b}^2 \he \rho_0 \dot R_{\rm b} = {2\lw\over v_{\rm w}}f(R_{\rm b},\he)(t-t_{\rm br})+{\pi \over 4}{4\pi\over 3}\left ({\he^7\lw\rho_0^2\over 17.9}\right)^{1/3}.
\end{equation}
\noindent
The term on the left hand side represents the shell momentum; the
first term on the right hand side describes the wind momentum
transferred to the shell after $t_{\rm br}$, while the second term
indicates the shell momentum at $t_{\rm br}${\footnote{\citet{koo92}
    considered only the shell momentum term; in fact, in their case,
    the first term is negligible because of the relatively low values
    of $L_{\rm w}$ considered.}.
\noindent
This equation can be rewritten as
\begin{equation}
\label{eq:eqrb2b}
\dot R_{\rm b,2}=A{f(R_{\rm b},\he)\over R_{\rm b,2}^2}(t_6-t_{\rm br,6})+{B\over R_{\rm b,2}^2},
\end{equation}
\noindent
where lengths are in 0.1 kpc and times in Myr. The two constants are
\begin{equation}
A=35.23{L_{41}\over  n_0 V_8},
\end{equation}
\noindent
and
\begin{equation}
B=2.09\left ({\hee^4L_{41}\over n_0}\right )^{1/3},
\end{equation}
\noindent
where $V_8\equiv V_{\rm w}/(10^8$ $\rm cm\;s^{-1}$), and $f(R_{\rm b},\he)$ is a geometrical factor evaluating 
the fraction of the total wind momentum in the plane which is transferred to the shell:
\begin{equation}
f(\theta)=\int_\theta^{\pi/2}\sin(\alpha)^2 {\rm d}\alpha=0.25\left (\pi-2\theta+\sin(2\theta)\right ),
\end{equation}
\noindent
where $\alpha$ is the angle between the normal to the plane and the radial wind velocity direction, 
and $\theta=\arctan(R_{\rm b}/H)$\footnote{At the end of the first stage it is $\theta \sim 0$, and $f(0)=\pi/4$;
this explains the presence of such numerical coefficient in the second term in the right end side of equation 
(\ref{eq:eqrb2}).}.

Equation (\ref{eq:eqrb2b}) can be integrated numerically. However, we find an excellent analytical approximation
(with a percentage error rarely above 3\%, and never higher than 6\%) to the numerical solution:
\begin{equation}
\label{eq:sana}
R(t)_{\rm b,2}=\hee\left (t\over t_{\rm br} \right )^\beta;
\end{equation}
\noindent
the exponent $\beta$ can be very well approximated by
\begin{equation}
\label{eq:esp}
\beta={\cal C}\left ({\cal L}\over 0.01\right )^\gamma,
\end{equation}
where ${\cal L}\equiv L_{41}/n_0$. The constant 
${\cal C}$ depends on $\he$: ${\cal C}=0.421$ for $\hee=1$, and ${\cal C}=0.397$ for $\hee=2$.
For $0.01\leq {\cal L}\leq 1$ it is $\gamma=0.039$ for both the values of $\he$; instead, if $0.001\leq {\cal L}<0.01$,
it is $\gamma=0.034$ for $\hee=1$, and $\gamma=0.017$ for $\hee=2$.

Equation (\ref{eq:sana}) holds up to $t_{\rm SN}$, the time at which
the SNe II stop to explode. At later times, however, the shell may
still keep expanding due to its ``inertia". Such an expansion is
regulated by an equation similar to equation \ref{eq:eqrb2}, but
without the contribution of the wind momentum
{ \footnote{During this stage, as during the previous ones, we neglect the ISM pressure during
the bubble expansion. The outward progress of the shell is thus maintained by the momentum
of the outward moving gas \citep{spitzer78}. The expansion stops at the dissolution of the shell
occurring when the shell velocity is about $\sigma_0$.}}
\begin{equation}
R_{\rm b}^2(t)\dot R_{\rm b}(t)=R_{\rm b}^2(t_{\rm SN})\dot R_{\rm b}(t_{\rm SN}).
\end{equation}
The above equation can be easily solved, and one obtains the time evolution of
the shell radius during this third stage:
\begin{equation}
R_{\rm b,2}(t)=\left (3\beta {R_{\rm b,2}^3(t_{\rm SN,6})\over t_{\rm SN,6}}(t_6-t_{\rm SN,6})+ R_{\rm b,2}^3(t_{\rm SN,6})\right )^{1/3}.
\end{equation}

In conclusion, we can summarize the evolution of the hole expansion as

%\[
 %\rh(t) = 
  %\begin{cases} 
   %R_1(t) & \text{if } t \leq t_{\rm br} \\
   %R_2(t)       & \text{if } t_{\rm br} <  t \leq t_{\rm SN} \\
   %R_3(t)       & \text{if } t>t_{\rm SN}
  %\end{cases}
%\]

\[
 R_{\rm b,2}(t_6) = 
  \begin{cases} 
   2.7{\cal L}^{1/5}t_6^{3/5} & \text{if } t \leq t_{\rm br} \\
   \hee\left (t\over t_{\rm br} \right )^\beta  & \text{if } t_{\rm br} <  t \leq t_{\rm SN} \\
   (3\beta {R_{\rm b,2}^3(t_{\rm SN,6})\over t_{\rm SN,6}}(t_6-t_{\rm SN,6}) + \\
   +  R_{\rm b,2}^3(t_{\rm SN,6}) )^{1/3}       & \text{if } t>t_{\rm SN}
  \end{cases}
\]

In the above equation we made use of the parameter ${\cal L}$ introduced earlier.
In fact, the wind luminosity and the density of the
ambient medium never occur separately in the equations, but always as
a ratio. Therefore the hole expansion is regulated only by two
parameters: ${\cal L}$ and, to a lesser extent, $\he$.

%%%%%%%%%%%%%%%%%%%%%%%%%%%%%%%%%%%%%%%%%%%%%%%%%%

% Don't change these lines
\bsp	% typesetting comment
\label{lastpage}
\end{document}